# PENGARUH PERANGKAT SERVER TERHADAP KUALITAS PENGONTROLAN JARAK JAUH MELALUI INTERNET

Gunawan[1)], Imam Muslim R[2)]
[1,2]Mahasiswa Magister (S2) Teknik Informatika
Universitas Sumatera Utara
Jl Universitas No 24 A Medan 20155 Indonesia
[1]gunawan@polmed.ac.id
[2]imamtkj@gmail.com

*Abstract*— Internet sangat membantu dalam memperbaiki kualitas kehidupan manusia. Hampir semua bidang kehidupan manusia dapat diakses menggunakan internet. Manusia terbantu dengan adanya internet yang *menyediakan* segala macam informasi yang dibutuhkan. Seiring dengan perkembangan infrastuktur jaringan internet maka pengontrolan jarak jauh mulai berubah menggunakan media internet. Pada penelitian ini digunakan server Notebook dan Raspberry Pi untuk mengetahui kualitas pengontrolan dari masing-masing perangkat server yang digunakan. Dalam penelitian ini akan diselidiki kemungkinan peningkatan kualitas remote control berbasis web dengan melakukan implementasi Raspberry Pi sebagai web server dan seberapa besar peningkatan kualitas remote control berbasis web yang diperoleh dalam penelitian ini.

Kata Kunci: Kontrol Jarak Jauh, Raspberry Pi, Server, Kualitas

## I. PENDAHULUAN

Internet sangat membantu masyarakat dalam memperbaiki kualitas kehidupan mereka. Hampir semua bidang kehidupan manusia dapat diakses menggunakan internet. Manusia terbantu dengan adanya internet yang menyediakan segala macam informasi yang mereka butuhkan.

Perkembangan teknologi internet erat kaitanya dengan penemuan-penemuan infrastruktur jaringan internet. Penemuan teknologi fiber optic dominan dalam peningkatan kecepatan akses internet. Dulu ketika masih menggunakan jaringan kabel telepon biasa, kecepatan akses internet hanya mencapai 28 Kbps. Sedangkan dengan fiber optic kecepatan internet meningkat sampai 100 Gbps diteliti oleh perusahaan Mitsubishi yang dipamerkan dalam ajang CIATEC (Cutting-Edge IT and Electronic Comprehensive Exhibition) pada tahun 2013.

Penelitian tentang ini terus berkembang mengikuti perkembangan teknologi di bidang informasi dan telekomunikasi. Dalam penelitian-penelitian sebelumnya pengontrolan jarak jauh (remote control) banyak memanfaatkan sinyal DTMF (Dual Tone Multi Frequency) dan SMS (Short Mesage Service) dari telepon seluler sebagai medianya. Sejauh ini pemanfaatan sinyal DTMF yang dikirim melalui telepon seluler dan dikodekan menjadi sinyal-sinyal digital untuk melakukan pengontrolan peralatan rumah tangga dan kantor (Das et al, 2009) masih menggunakan microcontroller sebagai hardware utamanya.

Penelitian yang sama juga dilakukan (Soufi et al, 2013). Mereka memanfaatkan sinyal DTMF sebagai media pengontrolan. Kemudian penelitian yang lainnya menggunakan SMS sebagai media pengontrolan yang berbasis teks dimana data berupa teks SMS dikodekan menjadi data untuk melakukan pengontrolan dari jarak jauh (Oke et al, 2013). Pada penelitian mereka sebagai server digunakan microcontroller untuk pusat kendali yang menjalankan semua perintah-perintah yang dikirim baik melalui DTMF maupun SMS dan perangkat modem sebagai media transmisi data DTMF maupun SMS.

Diagram blok pada gambar 1 menunjukkan prinsip kerja secara umum sistem pengontrolan jarak jauh berbasis DTMF dan SMS melalui media telepon seluler (Das et all, 2009).

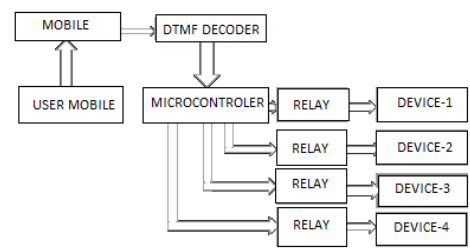

Gambar 1. Diagram Blok Sistem Pengontrolan Jarak Jauh Menggunakan Telepon Seluler.

Seiring dengan perkembangan infrastuktur jaringan internet maka pengontrolan jarak jauh mulai bergeser menggunakan media internet, dan penelitian tentang ini pun banyak dilakukan (Donghui Yu, 2012). Penelitian tentang pengontrolan jarak jauh melalui internet sudah menggunakan perangkat server komputer biasa (PC) namun masih memanfaatkan microcontroller sebagai interface agar dapat terhubung ke perangkat yang lain (Xiaowen,2013).

Diagram blok sistem pengontrolan jarak jauh menggunakan Zigbee (Hwang et al, 2003) ditunjukan pada gambar 1.2.

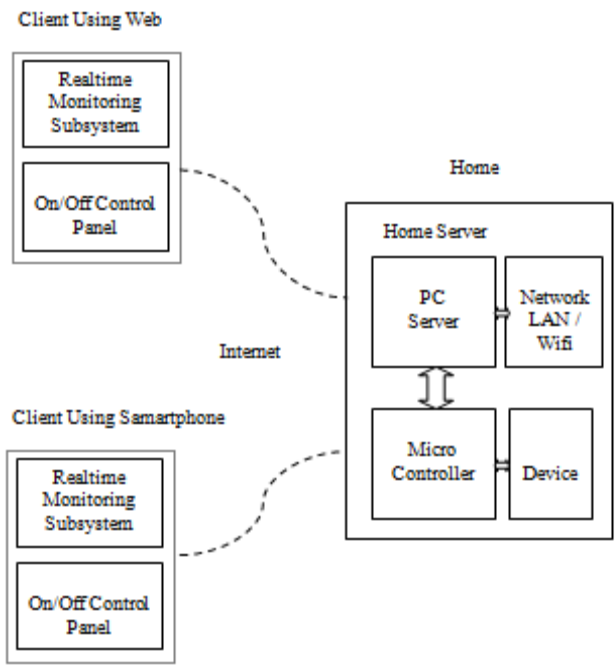

Gambar 2. Diagram Blok Sistem Pengontrolan Jarak Jauh Menggunakan Jaringan Internet

Hasil-hasil penelitian terdahulu dalam penerapannya di lapangan kelihatannya masih banyak mengalami kendala. Salah satu kendalanya adalah bahwa kualitas pengontrolan jarak jauh melalui internet ditentukan oleh kondisi perangkat server yang digunakan. Dengan penelitian ini kemungkinan kendala-kendala itu dapat diatasi. Salah satu caranya adalah dengan memanfaatkan perangkat Raspberry Pi sebagai server.

### A. Perumusan Masalah

Dalam penelitian ini akan diselidiki kemungkinan peningkatan kualitas remote control berbasis web dengan melakukan implementasi Raspberry Pi sebagai web server dan seberapa besar peningkatan kualitas remote control berbasis web yang diperoleh dalam penelitian ini.

### B. Batasan Masalah

Karena keterbatasan waktu, sumber daya, dana dan kemampuan keilmuan penulis, maka dalam penelitian ini hanya akan dilakukan hal-hal sebagai berikut:

a) Melakukan *uji coba dengan perangkat server berupa komputer mini yang dapat dihubungkan langsung dengan driver relay peralatan dan membutuhkan konsumsi daya listrik yang rendah.*
b) *Membatasi* pengukuran *kualitas hanya pada variabel kecepatan dan keberhasilan pengontrolan dalam beberapa kali sampel pengujian.*
c) Penelitian *menggunakan sumber daya berbasis open source yang tidak memerlukan pembiayaan.*

### C. Tujuan Penelitian

Penelitian ini bertujuan untuk:

a) Mengetahui kualitas perangkat server yang digunakan sebagai perangkat pengontrolan jarak jauh melalui internet.
b) Mengetahui peningkatan kualitas dengan melakukan implementasi Raspberry Pi sebagai web server untuk aplikasi remote control berbasis web.
c) Mendapatkan kualitas yang lebih baik dibandingkan dari hasil-hasil penelitian terdahulu.

### D. Manfaat Penelitian

Manfaat yang diharapkan dari penelitian ini adalah:

a) Pemanfaatan pada bidang Industri, implementasi hasil penelitian ini diharapkan dapat meningkatkan kualitas perangkat server yang berfungsi sebagai pengontrol peralatan sehingga mesin-mesin industri yang dikendalikan dari jarak jauh semakin efisien.
b) Pemanfaatan pada bidang pertahanan dan keamanan (militer), diharapkan hasil penelitian ini dapat diterapkan untuk pengendalian mesin-mesin perang dari jarak jauh dengan bentuk yang lebih kecil.
c) Bagi manajemen gedung perkantoran, perhotelan dan rumah tangga dapat memanfaatkan penelitian ini untuk melakukan pengontrolan dan pengamanan gedung, kantor dan rumah tinggal..

## II. LANDASAN TEORI

### A. Jaringan Komputer

Dalam penelitian ini jaringan internet diperlukan sebagai media transmisi data pengontrolan (*controled code*). Protokol yang digunakan adalah TCP IP dengan web server sebagai domain web pengendali. Interface jaringan internet yang digunakan adalah modul USB Wifi yang kompatibel dengan perangkat Raspberry server.

Seperti penelitian sebelumnya yang dilakukan oleh Soyoung Hwang dan Donghui Yu, mereka menggunakan Zig Bee Network sebagai modul Wifi dalam melakukan transmisi data pengontrol. Kelemahan penelitian Soyoung masih menggunakan server PC biasa sehingga membutuhkan interface perangkat microcontroller eksternal.

Dalam penelitian ini pengamanan web dilengkapi password dengan MD5 (Message Digest 5) sebagai metode enkripsi. Pemilihan metode ini disebabkan MD5 memiliki tingkat keamanan tinggi dan merupakan fasilitas dari pemrograman PHP yang *open source*.

Penelitan terdahulu banyak memanfaatkan media transmisi seluler dengan metode DTMF (*Dual Tone Multifrequency*) dan SMS (*Shot Message Service*) yang memiliki beberapa kelemahan antara lain:

a) *Transmisi data menggunakan metode voice yang rentan terhadap noise dan kualitas kabel transmisi.*
b) *Kecepatan pengiriman data voice relatif lambat dibanding dengan keceptan data internet.*
c) *Penggunaan terbatas hanya pada perangkat yang kompatibel saja.*
d) *Memerlukan proses dialing dan membutuhkan delay saat transmisi.*
e) *Jaringan voice dan SMS sangat tergantung dengan jaringan pada operator seluler yang digunakan.*
f) *Jaringan internet dapat diakses dari telepon seluler sehingga memungkinkan pengontrolan jarak jauh dari telepon seluler (smartphone) yang terkoneksi dengan jaringan internet. Dengan sistem ini maka perbedaan platform tidak menjadi permasalahan lagi.*

Pada penelitian ini penulis menggunakan jaringan internet tanpa kabel (wireless network). Jaringan ini menggunakan Wifi hotspot yang bersumber dari smartphone Samsung Galaxy Y Duos yang dapat diatur sebagai Wifi hotspot.

Dengan menggunakan jaringan Wifi maka diperoleh beberapa keuntungan yaitu:

a) Tidak membutuhkan kabel UTP yang panjang.
b) Tidak membutuhkan instalasi kabel yang relatif mahal.
c) Memanfaatkan wifi server notebook yang sudah ada
d) Pada Raspberry Pi perlu ditambah wireless network adapter yang menggunakan perangkat WNC0305USB.

### B. Komputer Mini Raspberry

Dalam penelitian ini penulis menggunakan komputer mini yang hanya sebesar kartu kredit. Komputer mini ini diproduksi oleh *Raspberry Foundation* yang berada di *United Kingdom* (UK). Komputer mini ini hanya membutuhkan daya listrik yang sangat kecil yaitu hanya 1,5 *Watt* untuk model A dan 3,5 *Watt* untuk model B. Sehingga cocok digunakan untuk pemakaian secara terus menerus.

Penelitian yang dilakukan penulis tidak membutuhkan microcontroller eksternal sebagai interface tambahan. Hal ini dimungkinkan karena dalam Raspberry telah terdapat port input output (GPIO). GPIO dengan jumlah 8 pin dapat melayani pengontrolan 8 buah perangkat yang berbeda.

### C. Sistem Operasi Linux Raspbian

Penelitian ini menggunakan sistem operasi *open source* yaitu Linux. Raspbian. Raspbian sudah dirancang khusus untuk pemakaian perangkat Raspberry dengan RAM 512 Mb. Kapasitas ini memang kecil dibanding yang dimiliki oleh perangkat server PC. Memory eksternal dapat diekspansi sampai 16GB dengan menggunakan SD Card eksternal yang dipasang pada SD Card Slot yang tersedia pada board Raspberry.

Dalam penelitian ini penulis memilih Raspbian karena mudah untuk diimplementasikan. Selain itu Raspbian mendukung percobaan dengan software pengontrolan jarak jauh menggunakan internet.

Raspbian sesuai digunakan dengan sistem ini dikarenakan dapat diinstall komponen-komponen yang diperlukan dalam pengontrolan peralatan listrik melalui web. Adapun software-software lain yang harus diinstall pada Raspbian ini adalah:

a) *Apache2 Web Server*
b) *MySQL untuk database server*
c) *PhpMyAdmin sebagai pengatur database*
d) *Control.php sebagai pengatur peralatan*
e) *GPIOServer.sh sebagai pengendali Pin Input Output Raspberry*

### III.  METODOLOGI PENELITIAN

#### A. Tahapan Penelitian

Pada penelitian ini, seluruh prosedur awal penelitian terlebih dahulu harus sudah dilakukan seperti studi literatur dan melakukan konsultasi dengan pembimbing. Setelah ditemukan permasalahan dan merumuskannya maka penelitian dapat dilanjutkan pada proses selanjutnya.

#### B. Teknik Pengembangan

Teknik pengembangan dalam penelitan ini mengikuti langkah-langkah sebagai berikut:

a) *Melakukan percobaan dengan menggunakan server komputer biasa. Pada tahap ini dilakukan pengujian menggunakan server komputer biasa yaitu Notebook Aspire One Happy. Melakukan uji coba pengiriman file dan pengambilan file (upload dan download).*
b) *Melakukan percobaan dengan menggunakan server komputer mini.Selanjutnya dilakukan pengujian sistem dengan menggunakan Raspberry Pi dengan sistem operasi Raspbian dan mengaplikasikan software yang telah dirancang. Kemudian melakukan proses pengiriman dan pengambilan file dari server Raspberry Pi.*
c) *Analisa dan evaluasi hasil.*
d) *Pada tahap ini dilakukan analisa terhadap hasil pengujian dan evaluasi kesalahan*

#### C. Tahapan Percobaan

Pada tahap ini dilakukan serangkaian percobaan dengan mengaplikasikan PC sebagai server. PC yang digunakan adalah Notebook dengan processor Intel Atom 1,67 GHz. Memory yang digunakan adalah 2 Gb. Sistem operasi yang digunakan adalah Windows 7 Ultimate berlisensi.

Langkah-langkah percobaan yang dilakukan adalah:

a) *Melakukan instalasi Apache web server dan Mysql yang akan berfungsi sebagai server web pengontrolan dari jarak jauh.*

b) *Membuat suatu program ujicoba yang berfungsi sebagai interface untuk server dan client. Program ini dibuat dengan bahasa pemrograman PHP.*

c) *Menjalankan program pada server dan melakukan setting agar dapat terjadi komunikasi antara client dan server.*

d) *Client menggunakan perangkan Android yaitu Smartphone Samsung Galaxy Y Duos.*

e) *Melakukan ujicoba system dan mengamati serta mencatat hasil-hasil pengukuran yang diperlukan.*

f) *Mengganti perangkat server dengan Raspberry Pi.*

g) *Melakukan instalasi webserver pada Raspberry Pi, tool yang digunakan adalah Apache2 dan Mysql.*

h) *Membuat program ujicoba dengan menggunakan bahasa pemrograman PHP.*

i) *Menjalankan program dan melakukan setting agar dapat terhubung dengan client.*

j) *Perangkat client yang digunakan adalah smartphone Samsung Galaxy Y Duos.*

k) *Melakukan ujicoba system dan mengamati serta mencatat hasil pengukuran yang dilakukan.*

D. Variabel yang diukur

Dalam penelitian ini ada beberapa variabel yang akan diukur, yaitu:

a) *Kecepatan pengiriman (upload) file dari client ke server*

b) *Kecepatan pengambilan file (download) dari server ke client.*

c) *Kecepatan respons server saat dikendalikan oleh client.*

Pada penelitian ini dilakukan pengujian pada dua buah server yaitu server Notebook Aspire One Happy menggunakan sistem operasi Windows 7 dan server komputer mini Raspberry Pi menggunakan sistem operasi Raspbian.

Pada penelitian ini dilakukan pengukuran kecepatan pengontrolan dengan menggunakan server komputer biasa dan membandingkannya dengan server Raspberry Pi. Diagram blok pengujian ditunjukkan pada gambar 3.

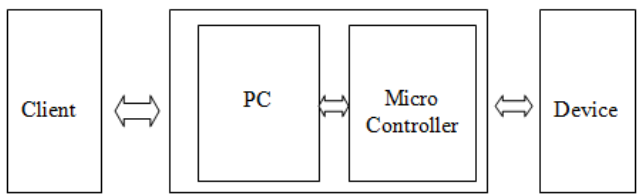

Gambar 3. Diagram Pengujian Menggunakan Server Komputer Notebook

Untuk diagram blok pengujian menggunakan server Raspberry Pi ditunjukkan pada gambar 4.

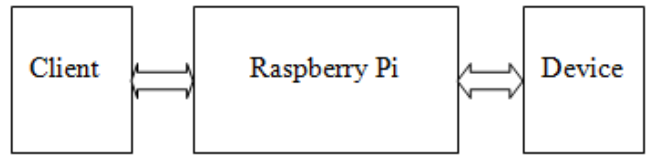

Gambar 4. Diagram Blok Pengujian Menggunakan Raspberry Pi

Sebelum melakukan pengujian maka hal-hal yang dilakukan untuk mempersiapkan server pada kedua server adalah sebagai berikut:

Persiapan Pada Server Notebook:
a) *Menginstall Aphace web server*
b) *Menginstall MySQL Server*
c) *Menginstall PhpMyAdmin*
d) *Menginstall FileZilla Server.*
e) *Membangun program WebCotroll.php menggunakan pemrograman PHP.*
f) *Melakukan instalasi driver USB4Rel untuk menghubungkan perangkat dengan Notebook.*

Persiapan pada Server Raspberry Pi:
a) *Menginstall Apche2 web server*
b) *Mengintall MySQL Server*
c) *Menginstall PhpMyAdmin*
d) *Membangun database untuk pengontrolan melalui web*
e) *Membangun program control.php*

IV. HASIL DAN PEMBAHASAN

Pada bagian ini akan dibahas hasil-hasil penelitian yang berkaitan dengan pemanfaatan komputer biasa yaitu Notebook Aspire One dan komputer mini Raspberry Pi sebagai server pengontrol peralatan listrik dari jarak jauh. Hasil penelitian dipaparkan dalam dua bagian yaitu hasil saat menggunakan server Notebook Aspire One dan Raspberry Pi. Kemudian akan dilakukan perbandingan hasil dari kedua server dan menganalisanya.

Penggunaan server yang berbeda ini dimaksudkan untuk mengetahui kelebihan dan kekurangan dari masing-masing server. Dengan demikian dapat diketahui server mana yang lebih cocok untuk digunakan sebagai server pengendali peralatan listrik dari jarak jauh melalui internet.

## A. Pengujian Pengontrolan Jarak Jauh Menggunakan Server Notebook

Tampilan program PHP untuk pengontrolan peralatan melalui internet menggunakan server notebook ditunjukkan pada gambar 4.

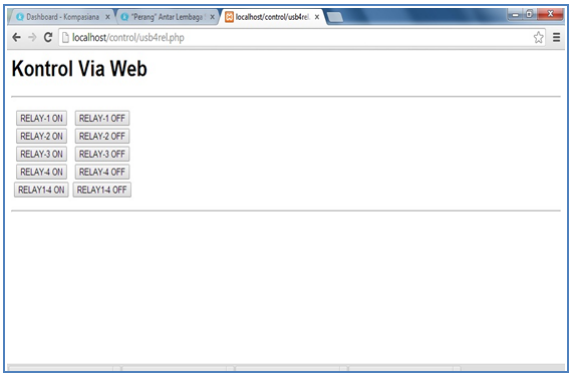

Gambar 5 Tampilan Program Kontrol Melalui Web

Sedangkan untuk pengujian pengontrolan jarak jauh penulis menggunakan perangkat Smartphone Samsung GT-S6102 dengan sistem Operasi Android Ginger Bread versi 2.3.6. Smartphone ini juga digunakan sebagai hotspot yang menyediakan jaringan Wifi untuk menghubungkan antara server dengan client.

Gambar 6 merupakan smartphone Samsung Galaxy Y Duos type GT-S6102 yang digunakan pada penelitian ini.

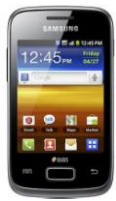

Gambar 6. Smartphone untuk pengujian sistem

Tampilan Program Kontrol via Web pada perangkat Android ditunjukkan pada gambar 7.

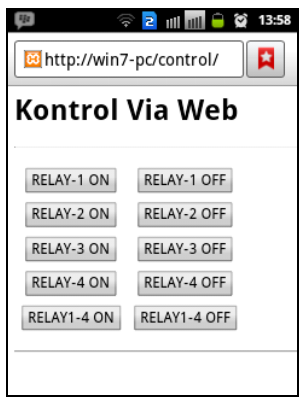

Gambar 7. Tampilan Program Web Kontrol pada Smartphone

Setelah perangkat terkoneksi dengan baik maka dilakukan ujicoba dengan menyentuh tombol pada perangkat Smartphone Android. Tabel pengujian ditunjukkan pada tabel 1.

Tabel 1. Pengujian Program Kontrol Melalui Web

| No | Tombol Control | Kondisi Led/Alat | | | | Response Time (det) |
|---|---|---|---|---|---|---|
| | | Led 1 | Led 2 | Led3 | Led 4 | |
| 1 | RELAY1-4 OFF | OFF | OFF | OFF | OFF | 0,132 |
| 2 | RELAY-1 ON | ON | OFF | OFF | OFF | 0.129 |
| 3 | RELAY-2 ON | ON | ON | OFF | OFF | 0.172 |
| 4 | RELAY-3 ON | ON | ON | ON | OFF | 0,153 |
| 5 | RELAY-4 ON | ON | ON | ON | ON | 0,126 |
| 6 | RELAY-1 OFF | OFF | ON | ON | ON | 0,121 |
| 7 | RELAY-2 OFF | OFF | OFF | ON | ON | 0,123 |
| 8 | RELAY-3 OFF | OFF | OFF | OFF | ON | 0,129 |
| 9 | RELAY-4 OFF | OFF | OFF | OFF | OFF | 0,124 |
| 10 | RELAY1-4 ON | ON | ON | ON | ON | 0,137 |

## B. Pengujian Kecepatan Server

Untuk pengukuran kecepatan transfer pengiriman file maka dilakukan uji coba dengan melakukan pengiriman file ke server (upload) dan sebaliknya dilakukan juga pengambilan file dari server (download). Proses pengukuran ini dilakukan dengan bantuan sebuah software yaitu Filezilla.

Filezilla adalah software yang berfungsi untuk melakukan transfer file melaui protokol FTP (File Transfer Protocol). FTP ini memang khusus dirancang agar masing-masing perangkat komputer dapat saling mengirim dan menerima file. Tool Filezilla ditunjukkan pada gambar 8.

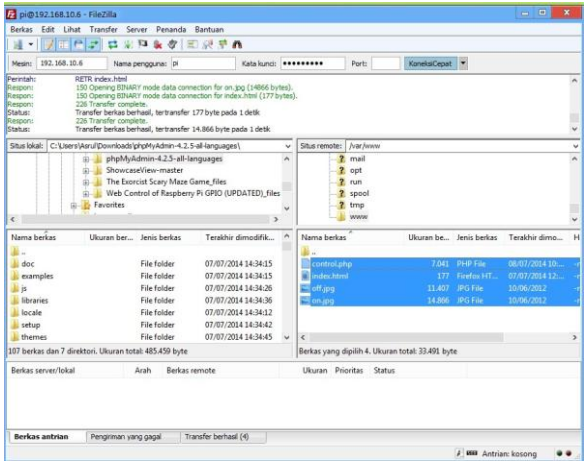

Gambar 8. Tampilan Tool Filezilla

Untuk melakukan transfer file, maka server Filezilla harus diaktifkan. Caranya adalah dengan mengaktifkannya melalui XAMPP Control dan kemudian mengatur user atau client yang dapat terhubung ke server. Gambar 9 menunjukkan XAMPP Control dengan Filezilla server yang telah diaktifkan.

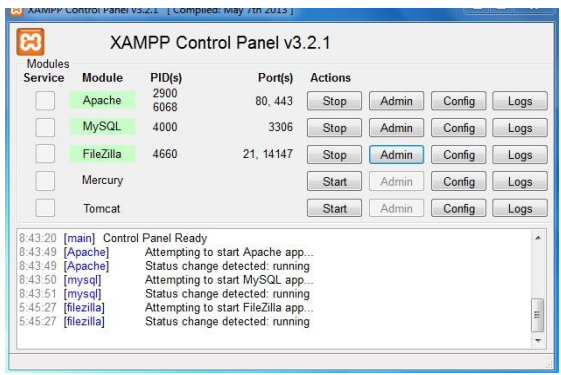

Gambar 9. Server FileZilla yang Sudah Diaktifkan.

Server FileZilla ini harus disetting dan dibuat user baru agar dapat terhubung dengan client. Dalam penelitian ini dibuat sebuah client dan dilindungi dengan sebuah password sehingga hanya user yang berhak saja yang boleh melakukan pengiriman dan pengambilan file.

### C. Pengujian Pengiriman File

Pada pengujian pengiriman file ini bertujuan untuk mengetahui kecepatan pengiriman file pada server notebook yang digunakan. Sebelumnya dipersiapkan 10 buah file dengan jumlah bit yang berbeda-beda. Kemudian file-file itu diunggah ke server menggunakan tools FileZilla.

Pada gambar 10. diperlihatkan saat dilakukan pengujian transfer file menggunakan FileZilla.

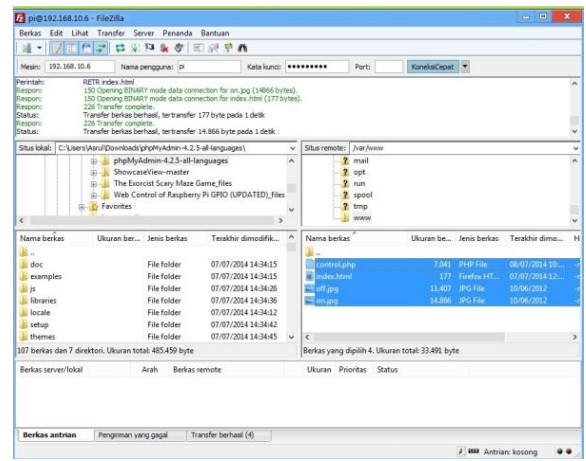

Gambar 10. Screenshoot saat melakukan proses transfer file dengan FileZilla.

Hasil pengukuran pengiriman file ditunjukkan pada tabel 2. File dikirim dari jaringan Wifi local.

Tabel 2. Pengukuran Waktu Pengiriman File

| No | Nama File | Ukuran (bit) | Waktu (detik) |
|---|---|---|---|
| 1 | MMD X.exe | 1.028.608 | 1 |
| 2 | NMDVPN.zip | 2.057.157 | 1 |
| 3 | Yii-docs.zip | 5.133.695 | 5 |
| 4 | Bluestack.exe | 10.245.808 | 11 |
| 5 | Spongebob.mp4 | 15.814.424 | 13 |
| 6 | Imto ipad.zip | 20.781.591 | 18 |
| 7 | Ipad2pc.zip | 31.838.979 | 31 |
| 8 | Ext407.zip | 41.665.162 | 38 |
| 9 | Anoying.mp4 | 51.452.516 | 47 |
| 10 | Counter Strike.zip | 92.274.688 | 87 |
| JUMLAH | | 272.292.628 | 252 |

Dari tabel 2 dapat dihitung kecepatan rata-rata pengiriman file dari server ke client dengan persamaan:

$$V = Jlh. Bit / waktu$$
$$V = 272292,62/252$$
$$V = 1.080.526,3 \text{ bit/detik}$$

### D. Pengujian Pengambilan File

Selanjutnya untuk uji pengambilan file (download) ditampilkan pada tabel 3. Pada tabel itu ditunjukkan bahwa hasil untuk proses download dan upload tidak menunjukkan perbedaan yang berarti. Hal ini disebabkan jaringan yang digunakan adalah jaringan Wifi local.

Tabel 3. Pengukuran waktu pengambilan file (download)

| No | Nama File | Ukuran (bit) | Waktu (detik) |
|---|---|---|---|
| 1 | MMD X.exe | 1.028.608 | 1 |
| 2 | NMDVPN.zip | 2.057.157 | 2 |
| 3 | Yii-docs.zip | 5.133.695 | 5 |
| 4 | Bluestack.exe | 10.245.808 | 9 |
| 5 | Spongebob.mp4 | 15.814.424 | 14 |
| 6 | Imto ipad.zip | 20.781.591 | 18 |
| 7 | Ipad2pc.zip | 31.838.979 | 31 |
| 8 | Ext407.zip | 41.665.162 | 38 |
| 9 | Anoying.mp4 | 51.452.516 | 42 |
| 10 | Counter Strike.zip | 92.274.688 | 83 |
| JUMLAH | | 272.292.628 | 243 |

Dengan cara yang sama data pada tabel 4.3 dapat dihitung kecepatan rata-rata pengambilan file dari server ke client dengan persamaan:

$$V = Jlh.Bit/waktu$$
$$V = 272292628/243$$
$$V = 1120545,79 \text{ bit/detik}$$

Hasil pengukuran ini diketahui bahwa untuk pengambilan file dari client ke server lebih cepat ketimbang mengirim file dari server ke client. Hal ini ditunjukkan bahwa terjadi selisih 40019,49 bit setiap detik.

*E. Pengujian Pengontrolan Jarak Jauh Menggunakan Server Raspberry Pi*

Setelah perangkat Raspberry Pi dihubungkan dengan keyboard, mouse, USB Wifi dan catu daya, maka Raspberry Pi aktif dengan menjalankan sistem operasinya yaitu Raspbian. Tampilan awal Raspbian ditunjukkan pada gambar 11.

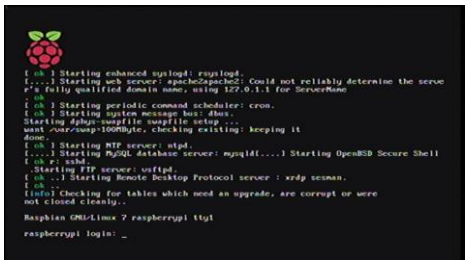

Gambar 11 Tampilan Raspbian Operating System

Untuk memulai menjalankan Raspbian harus login terlebih dahulu dengan user Pi dan passwordnya adalah raspberry. Tampilan Raspbian OS pada mode grafik ditunjukkan pada gambar 12.

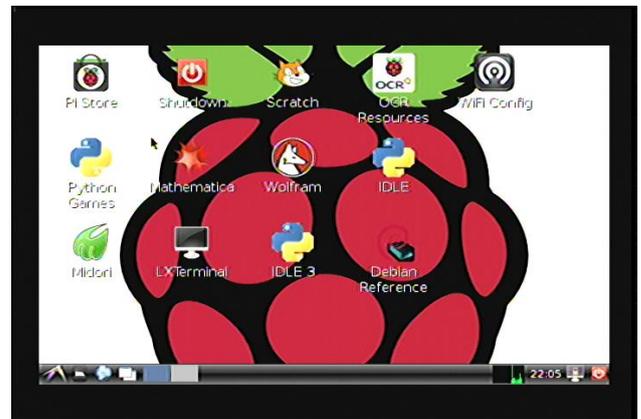

Gambar 12. Tampilan Raspbian OS dalam Mode Grafik

Uji coba pengontrolan dari web pada server Raspberry Pi dilakukan dengan perangkat Laptop Acer dan Smartphone Samsung GT-S6102. Untuk tampilan web control yang dibuat dengan php ditunjukkan pada gambar 13.

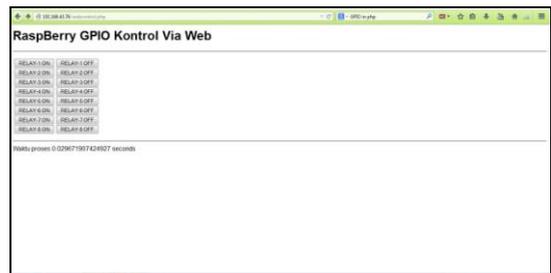

Gambar 13. Tampilan Web Control untuk GPIO Raspberry Pi

Setelah dilakukan pengontrolan maka pada Raspberry dapat diketahui kondisi Pin GPIO yang On dan Off sesuai dengan pengaturan yang dilakukan melalui web. Pengontrolan melalui web ini juga dapat dilakukan melalui Smartphone, dalam hal ini pengujian dilakukan dengan Smartphone Samsung GT-S61102 dengan sistem operasi Android Ginger Bread. Tampilan pengontrolan melalui Smartphone ditunjukkan pada gambar 14.

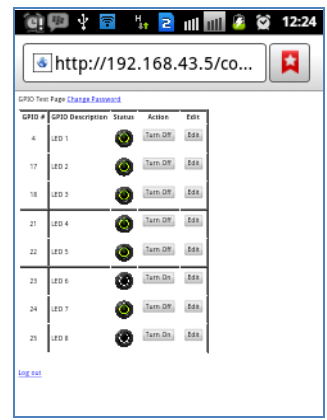

Gambar 14. Tampilan Control.php pada Smartphone Samsung GT-S61102

Setelah program Control.php berjalana, maka untuk keamanan diminta login dengan username "admin" dan password "mygpio45". Kemudian dapat mengklik atau menyentuh tombol pada kolom action. Jika kondisinya sedang on maka setalah diklik akan berubah menjadi off. Hasil program ditunjukkan pada gambar 15.

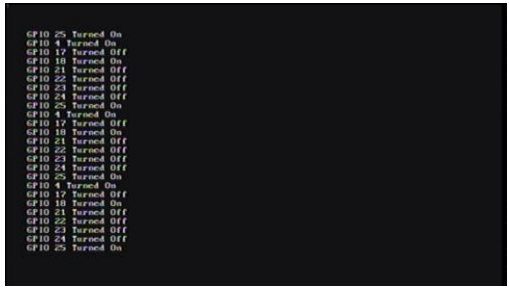

Gambar 15. Hasil pada server saat program control.php dijalankan.

Tampilan pada gambar 4.11 adalah hasil dari pembacaan database kondisi Pin GPIO yang tersimpan dalam server. Untuk mengaksesnya maka terlebih dahulu harus menjalankan script ./GPIOServer.sh pada server Raspberry Pi. Hasil pengujian ini lebih lengkapnya ditunjukkan pada tabel 4.5.

Tabel 5. Pengujian Program kontrol via Web

| No.PIN | LED/Alat | Action Button | Kondisi LED/Alat | Response Time (det) |
|---|---|---|---|---|
| 4 | LED 1 | Turn Off | On | 0,078 |
| 17 | LED 2 | Turn On | Off | 0,075 |
| 18 | LED 3 | Turn Off | On | 0,051 |
| 21 | LED 4 | Turn On | Off | 0,055 |
| 22 | LED 5 | Turn On | Off | 0,07 |
| 23 | LED 6 | Turn On | Off | 0,064 |
| 24 | LED 7 | Turn On | Off | 0,066 |
| 25 | LED 8 | Turn Off | On | 0,078 |

*F. Pengujian Kecepatan Server Raspberry Pi*

Raspberry Pi dapat diukur kecepatan transmisi file dengan terlebih dahulu memasang vsftpd agar dapat berkomunikasi dengan komputer lain menggunakan tools FileZilla. Pada penelitian ini digunakan juga 10 buah file yang sama dengan file-file yang diuji coba pada server notebook.

Proses transfer file pada server Raspberry Pi menggunakan software FileZilla ditunjukkan pada gambar 16.

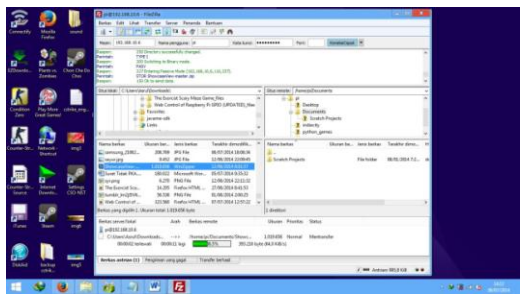

Gambar 16 Proses transfer File pada server Raspberry Pi

*G. Pengujian Pengiriman File*

Client yang digunakan untuk pengujian ini adalah laptop Acer dengan processor Pentium Centrino dan memiliki memory RAM 4 GB. Dengan melakukan pengiriman 10 buah file ke server Rasperry Pi dan mencatat waktu pengirimannya, maka hasilnya diperlihatkan pada tabel 6.

Tabel 6. Pengukuran waktu pengiriman file

| No | Nama File | Ukuran (bit) | Waktu (detik) |
|---|---|---|---|
| 1 | MMD X.exe | 1.028.608 | 9 |
| 2 | NMDVPN.zip | 2.057.157 | 16 |
| 3 | Yii-docs.zip | 5.133.695 | 39 |
| 4 | Bluestack.exe | 10.245.808 | 77 |
| 5 | Spongebob.mp4 | 15.814.424 | 136 |
| 6 | Imto ipad.zip | 20.781.591 | 154 |
| 7 | Ipad2pc.zip | 31.838.979 | 233 |
| 8 | Ext407.zip | 41.665.162 | 301 |
| 9 | Anoying.mp4 | 51.452.516 | 404 |
| 10 | Counter Strike.zip | 92.274.688 | 738 |
| JUMLAH | | 272.292.628 | 2107 |

Dari data pada tabel 4.6 maka dapat dihitung kecepatan rata-rata pengiriman file dari server ke client dengan persamaan:

$$V = Jlh.Bit/waktu$$
$$V = 272292628/2107$$
$$V = 129.232,4 \text{ bit/detik}$$

*H. Pengujian Pengambilan File*

Dengan cara yang sama seperti pada server notebook, maka untuk pengujian pengambilan file (download) hasilnya ditampilkan pada tabel 7.

Tabel 7. Pengukuran waktu pengambilan file (download)

| No | Nama File | Ukuran (bit) | Waktu (detik) |
|---|---|---|---|
| 1 | MMD X.exe | 1.028.608 | 5 |
| 2 | NMDVPN.zip | 2.057.157 | 10 |
| 3 | Yii-docs.zip | 5.133.695 | 25 |
| 4 | Bluestack.exe | 10.245.808 | 52 |
| 5 | Spongebob.mp4 | 15.814.424 | 71 |
| 6 | Imto ipad.zip | 20.781.591 | 96 |
| 7 | Ipad2pc.zip | 31.838.979 | 164 |
| 8 | Ext407.zip | 41.665.162 | 186 |
| 9 | Anoying.mp4 | 51.452.516 | 232 |
| 10 | Counter Strike.zip | 92.274.688 | 411 |
| JUMLAH | | 272.292.628 | 1252 |

Dengan cara yang sama data pada tabel 7 dapat dihitung kecepatan rata-rata pengambilan file dari server ke client dengan persamaan:

$$V = Jlh.Bit/Waktu$$
$$V = 272.292.628/1252$$
$$V = 217.486{,}1 \text{ bit/detik}$$

*I. Perbandingan Hasil*

Dari data-data hasil penelitian yang telah diperoleh dapat dilakukan perbandingan bahwa dapat diketahui kecepatan transfer data lebih cepat pada server yang menggunakan notebook dibanding dengan server Raspberry Pi.

Dari hasil pengujian dapat dibandingkan antara kedua server dengan data pada tabel 8 berikut ini.

Tabel 8 Perbandingan kecepatan transfer File pada server

| No | Server | Upload (bit/s) | Download (bit/s) |
|---|---|---|---|
| 1 | Notebook | 1.080.526 | 1.120.546 |
| 2 | Raspberry Pi | 129.232 | 217.486 |
| Rasio Kecepatan | | 8 | 5 |

Hasil perbandingan pada tabel 4.8 dapat digambarkan pada grafik yang ditunjukkan pada gambar 17.

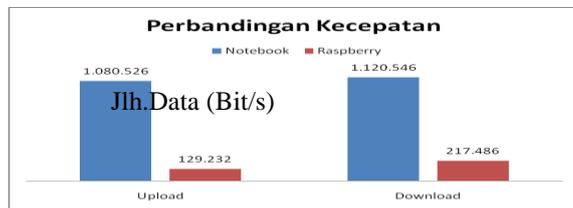

Gambar 17. Grafik Perbandingan Kecepatan Transfer data pada Server

Dari hasil penelitian ini dapat diketahui bahwa server Raspberry Pi lebih lambat dalam hal untuk trasfer data file dengan jumlah yang besar. Secara spesifikasi memang kecepatan prosesor dan arsitektur sangat mempengaruhi dalam hal kecepatan pengiriman data.

Untuk tampilan grafik line perbandingan kecepatan transfer data antara server notebook Aspire One Happy dengan server Raspberry Pi ditunjukkan pada gambar 18.

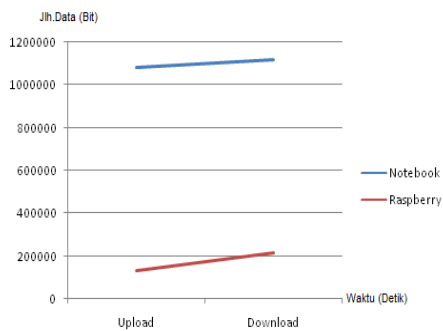

Gambar 18. Perbandingan Kecepatan Server Notebook dan Raspberry Pi

Penggunaan Raspberry cocok sebagai server pengontrolan peralatan yang tidak membutuhkan transmisi data yang besar. Dalam uji coba diperoleh response time yang hampir sama dari server notebook dan Raspberry Pi. Sehingga penggunaan Raspberry sebagai server pengontrolan peralatan tidak mengalami kendala.

V. KESIMPULAN

Dari hasil penelitian dan pembahasannya maka dapat diambil beberapa kesimpulan:

1. Server yang menggunakan Notebook harus menggunakan perangkat Microcontroller agar dapat digunakan sebagai perangkat server untuk pengontrolan peralatan dari jarak jauh.
2. Raspberry Pi tidak membutuhkan perangkat tambahan berupa microcontroller sebagai pengendali perangkat lain karena Raspberry Pi sudah dilengkapi Pin GPIO yang dapat dimanfaatkan untuk keperluan pengontrolan peralatan listrik melalui web.
3. Raspberry Pi tidak cocok digunakan untuk server yang memerlukan penyimpanan data yang besar (hosting file).
4. Untuk penggunaan yang tidak memerlukan transfer data yang besar Raspberry Pi masih memenuhi syarat dan tidak mengalami kendala.

## *Daftar Pustaka*